\documentstyle[12pt,twoside]{article}
\input psfig.sty
\def\double12 {\smallskipamount=6pt plus2pt minus2pt
                  \medskipamount=12pt plus4pt minus4pt
                  \bigskipamount=24pt plus8pt minus8pt
                  \normalbaselineskip=24pt plus0pt minus0pt
                  \normallineskip=2pt
                  \normallineskiplimit=0pt
                  \jot=6pt
                  {\def\smallskip {\vskip\smallskipamount}}
                  {\def\medskip   {\vskip\medskipamount}}
                  {\def\bigskip   {\vskip\bigskipamount}}
                  {\setbox\strutbox=\hbox{\vrule
                    height17.0pt depth7.0pt width 0pt}}
                  \parskip 0pt
                  \normalbaselines}

\def\half12 {\smallskipamount=6pt plus2pt minus2pt
                  \medskipamount=12pt plus4pt minus4pt
                  \bigskipamount=24pt plus8pt minus8pt
                  \normalbaselineskip=16pt plus0pt minus0pt
                  \normallineskip=2pt
                  \normallineskiplimit=0pt
                  \jot=6pt
                  {\def\smallskip {\vskip\smallskipamount}}
                  {\def\medskip   {\vskip\medskipamount}}
                  {\def\bigskip   {\vskip\bigskipamount}}
                  {\setbox\strutbox=\hbox{\vrule
                    height17.0pt depth7.0pt width 0pt}}
                  \parskip 0pt
                  \normalbaselines}

\def\single12 {\smallskipamount=3pt plus2pt minus2pt
                  \medskipamount=6pt plus4pt minus4pt
                  \bigskipamount=12pt plus8pt minus8pt
                  \normalbaselineskip=12pt plus0pt minus0pt
                  \normallineskip=1pt
                  \normallineskiplimit=0pt
                  \jot=3pt
                  {\def\smallskip {\vskip\smallskipamount}}
                  {\def\medskip   {\vskip\medskipamount}}
                  {\def\bigskip   {\vskip\bigskipamount}}
                  {\setbox\strutbox=\hbox{\vrule
                    height8.5pt depth3.5pt width 0pt}}
                  \parskip 0pt
                  \normalbaselines}

\def\refitem{\par\noindent\hangindent 20pt}

\def\wisk#1{\ifmmode{#1}\else{$#1$}\fi}

\def\deg    {\wisk{^\circ}}
\def\ddeg   {\wisk{{\rlap.}^\circ}}

\begin{document}
\pagestyle{plain}
\half12

\large
\begin{center}
Simulations of Foreground Effects for CMB Polarization
\end{center}

\medskip
\normalsize
\half12
\noindent
\begin{center}
A.~Kogut\footnotemark[1]$^{,2}$
and
G. Hinshaw$^1$
\end{center}
\footnotetext[1]{
Laboratory for Astronomy and Solar Physics, 
Code 685, NASA/GSFC, Greenbelt MD 20771. \newline
\indent~$^2$ E-mail: Alan.Kogut.1@gsfc.nasa.gov. \newline
}

\medskip
\normalsize
\half12
\begin{center}
Accepted for publication in \\
{\it The Astrophysical Journal} \\
\end{center}


\medskip
\begin{center}
\large
ABSTRACT
\end{center}

\normalsize
\noindent
We use a simple model to investigate the effect 
of polarized Galactic foreground emission
on the ability of planned CMB missions
to detect and model CMB polarization.
Emission from likely polarized sources
(synchrotron and spinning dust)
would dominate the polarization of the microwave sky
at frequencies below 90 GHz
if known Galactic foregrounds are at least 10\% polarized 
at high latitude ($|b| > 30\deg$).
Maps of polarization
at frequencies below 90 GHz
will likely require correction for foreground emission
to enable statistical analysis 
of the individual Stokes Q or U components.
The temperature-polarization cross-correlation
is less affected by foreground emission,
even if the foreground polarization is highly correlated
with the foreground intensity.
Polarized foregrounds, even if uncorrected,
do not dominate the uncertainty in the
temperature-polarization cross-correlation
for instrument noise levels typical of the MAP experiment.
Methods which remove galactic signals at the cost of
signal to noise ratio should carefully balance the
value of rejecting faint foregrounds
with the cost of increased instrument noise.

\clearpage
\normalsize
\section{Introduction}
Polarization of the cosmic microwave background (CMB)
provides a powerful probe of physical conditions and processes
in the early universe.
Following the detection of CMB temperature anisotropy,
a great deal of theoretical effort has gone into
calculating the polarized signal in various cosmological models
(for a recent review, see
Hu \& White 1997).
Precise maps of the polarization of the cosmic microwave background
complement maps of the temperature anisotropy
and offer the opportunity to break 
the parameter degeneracy
that occurs in fitting temperature data alone
(Zaldarriaga, Spergel, \& Seljak, 1997;
Efstathiou, \& Bond, 1999;
Kinney 1998).

CMB polarization is generated by Thomson scattering at recombination
and is typically a small fraction of the temperature anisotropy.
At expected amplitudes of only a few $\mu$K,
polarized emission of cosmic origin
could easily be masked by polarized foreground emission
from the Galaxy.
Little is known about the large-scale polarization
of high-latitude Galactic emission.
Synchrotron emission from electrons accelerated in the Galactic magnetic field 
is the dominant foreground at frequencies below $\sim$20 GHz
and is known to be linearly polarized.
Extrapolation of radio polarization maps
(Brouw \& Spoelstra 1976)
to millimeter wavelengths
indicate a polarization fraction between 10\% and 75\%
depending on Galactic latitude
(Lubin \& Smoot  1981).
Free-free emission from electron-ion collisions
in the interstellar medium
was thought to be the brightest foreground at millimeter wavelengths,
but remains largely un-mapped.
Although free-free emission is expected to be unpolarized
(except for small contributions from secondary Thomson scattering),
recent work hints that additional microwave emission
associated with infrared cirrus
may have been mis-identified as free-free emission
(Kogut et al.\ 1996, 1997;
Leitch et al.\ 1997;
de Oliveira-Costa et al.\ 1998, 1999).
Radiation from spinning dust grains has been suggested as a
possible mechanism for this anomalous emission
(Ferrara \& Dettmar 1994;
Draine \& Lazarian 1998, 1999).
Thermal emission from vibrational modes in the grains
should contribute negligibly to polarization
at frequencies below 90 GHz,
but electric dipole radiation from spinning dust 
is expected to be polarized
as the dust grains align with the Galactic magnetic field,
and could be a significant foreground
at frequencies between 20 and 90 GHz.

Galactic emission is bright enough
at microwave frequencies
that even partial polarization
could mask signals of cosmological interest.
Several authors have investigated techniques to
recover the cosmic signal in the presence of polarized foregrounds
(Bouchet, Prunet, \& Sethi 1999;
Tegmark et al.\ 2000).
These techniques use the average frequency and angular dependence
of foreground emission to construct minimum-variance CMB maps 
subject to the constraint that 
emission matching these foreground dependences be cancelled.
Although correct in principle,
the foreground cancellation comes at a price in sensitivity.
The linear combination of frequency channels
designed to suppress foreground emission
is weighted by the frequency dependence of the cosmic and foreground signals
and is significantly noisier than a 
simple noise-weighted sum.
Bouchet et al.\ (1999) 
conclude that the forthcoming MAP mission,
corrected in this fashion,
will have only 
marginal sensitivity to CMB polarization
in the presence of polarized foregrounds.

Foreground removal involves a tradeoff 
between instrument noise and foreground residuals; 
the optimum strategy depends on both quantities. 
Methods appropriate in the limit of bright contaminating foregrounds
may not be necessary if the foregrounds are sufficiently faint.
We analyze simulated maps
on 0\ddeg2 angular scales
including models of polarized Galactic emission
to determine whether the polarized foregrounds
require detailed models of the emission sources,
or if the effect is weak enough that
simpler, more sensitive remedies can be employed.

\section{Polarized Galactic Model}

Galactic polarization at millimeter wavelengths is not known
in any detail.
Known foregrounds, though,
do not closely mimic the specific polarization pattern
expected for the CMB signal,
suggesting that a toy model capturing the approximate amplitude
and pattern of foreground polarization
is sufficient.
For example, the cosmological temperature-polarization cross-correlation
is characterized by a bulls-eye pattern surrounding hot and cold spots.
Foreground polarization, on the other hand, 
reflects the orientation of the Galactic magnetic field
projected on the plane of the sky.
In order for foregrounds to produce a bulls-eye, the magnetic field
would have to point radially outwards from a single point.
Such a pattern is inconsistent with measurements of the Galactic magnetic
field inferred from either synchrotron or scattered dust polarization
(Brouw \& Spoelstra 1976;
Mathewson \& Ford 1970).
Our simulations thus use 
a simple toy model of polarized Galactic emission
intended to capture the important features
of astrophysically plausible polarization 
without assuming detailed knowledge of the 
interstellar medium at high latitudes.

We adopt the empirical ``correlation'' COBE-DMR model 
(Hinshaw et al.\ 1996 Table 1)
for the unpolarized Galactic intensity (Stokes I component),
scaling the Haslam 408 MHz survey 
(Haslam et al.\ 1981)
to model synchrotron emission, 
and scaling the DIRBE 100 $\mu$m survey for 
both the dust and free-free (or spinning dust) components.
We assume that the polarized Galactic emission is proportional
to the unpolarized intensity, and model the Stokes Q and U components as
\begin{eqnarray}
Q = f \cos(2\gamma) ~I \nonumber \\
U = f \sin(2\gamma) ~I,
\label{pol_def}
\end{eqnarray}
where
$f(l,b)$ is the fractional polarization,
assumed to vary across the sky,
and $\gamma(l,b)$ is the polarization angle
defined with respect to 
meridians connecting the Galactic poles.
Although we do not expect this to be true in detail,
it provides a simple way to model polarized emission 
while retaining the non-Gaussian features typical of
diffuse Galactic emission processes.

We wish to determine whether Galactic contamination 
of a simple noise-weighted CMB map produces a significant degradation
compared to the instrument noise.
A detailed approach would 
produce separate Q and U maps for each Galactic emission component
in each frequency channel,
and then co-add the channels weighted by the instrument noise.
For simplicity, 
we instead evaluate the various Galactic emission components 
at a single fiducial frequency,
and then scale this multi-component intensity map
to produce maps of the foreground Q and U polarization.
We adopt fiducial frequency 40 GHz
(the lowest the three co-added MAP channels)
with mean fractional polarization $\langle f \rangle = 0.1$.
The RMS amplitude of the resulting foreground Q or U fluctuations 
is 3.2 $\mu$K.
The choice of fiducial frequency and mean polarization fraction
reflects a conservative estimate of likely foreground contributions
averaged over the MAP channels at 40, 60, and 90 GHz.
A pessimistic model with 70\% synchrotron polarization
and 10\% dust polarization would produce
RMS fluctuations of 3.0 $\mu$K in the Stokes Q or U components
after co-adding the individual channel maps.
Both synchrotron and spinning dust polarization, 
when averaged over the high-latitude sky,
are likely to have smaller fractional polarization 
then this pessimistic case.
The toy model thus produces a simple, conservative estimate
for the amplitude and spatial distribution of foreground polarization.

At high latitudes, 
the foreground polarization angle $\gamma$
tends to follow the known radio loops,
showing coherent structure 
on the angular scales $\theta < 20\deg$ 
relevant to the CMB signal
(Mathewson \& Ford 1970;
Brouw \& Spoelstra 1976).
We approximate this behavior in our toy model
by generating simulated fields $\gamma(l,b)$ 
with different coherence angles $\theta_c$.
We first first populate each pixel with 
random $[x,y]$ coordinates
drawn independently from a uniform distribution.
We normalize each $[x,y]$ pair to unit amplitude in each pixel,
and then smooth the resulting maps
with Gaussian full width at half maximum 
$\theta_c = [0, 2, 5, 10, 30]$ degrees, respectively.
After smoothing, we generate a new normalization $n$
for each coordinate pair 
and assign $\gamma = \arctan(y,x)$ in the usual fashion.
Smoothing a vector field on the surface of a sphere
produces topological defects
where the angle $\gamma$ becomes undefined
(the normalization $n \rightarrow 0$).
To prevent these topological ``kinks'' in the smoothed fields
from dominating the small-scale distribution of the Q and U components,
we use the smoothed normalization $n$
to define a spatially-varying polarization amplitude,
\begin{equation}
f(l,b) = a ~n(l,b)
\label{norm_eq}
\end{equation}
to force the polarized emission to zero at each defect.
The scale factor $a$ is chosen so that the mean polarization
fraction $\langle f \rangle = 0.1$
averaged over the full sky,
compatible with the 
polarization fraction expected from the dominant foregrounds
(synchrotron or spinning dust)
at frequencies below 100 GHz.

Since our analysis involves only rotationally-invariant quantities
(Eq. \ref{corr_def}),
the result can not depend on the 
coordinate-dependent orientation of $\gamma$ in any one region,
but only on the {\it change} in orientation across that region. 
We employ the coherence angle to assure that 
the orientation in our toy model
remains nearly constant on scales $\theta < \theta_c$.
By varying the value of $\theta_c$ we investigate
the dependence of our results 
on the local coherence of the polarization angle,
ranging from incoherent ($\theta_c = 0$)
to coherent over large regions ($\theta_c = 30\deg$).

\section{Simulations}

We use simulated maps of cosmic and Galactic emission
to examine the effects of polarized Galactic foregrounds
on the ability to detect a cosmic polarization signal.
For clarity, we specialize to the parameters of the planned MAP mission
(Bennett et al.\ 1995),
although the results are applicable to a broad range of CMB missions.
We use the CMBFAST power spectra
(Seljak \& Zaldarriaga 1996)
smoothed with a 0\ddeg2 beam
and generate random realizations 
of CMB sky maps 
in Stokes I, Q, and U components,
including the proper correlation
between the temperature and polarization.
For the purposes of this paper,
we restrict analysis to two input models:
standard CDM 
and a reionized model (optical depth $\tau = 0.5$)
with significant large-scale polarization.
Both models have degree-scale polarization
produced at the recombination epoch;
reionization adds additional polarized emission
on scales of tens of degrees.
To each realization we then add Galactic emission (Eq. \ref{pol_def})
and random realizations of instrument noise.
The mean noise per $0\ddeg2 \times 0\ddeg2$ pixel is
27 $\mu$K for Stokes I 
and 37 $\mu$K for Q or U,
consistent with the noise properties
of the co-added MAP data in the 40, 60, and 90 GHz channels
over a 2-year mission.
There is negligible correlation between the noise
in the intensity and polarization maps.

\begin{figure}[b]
\centerline{
\psfig{file=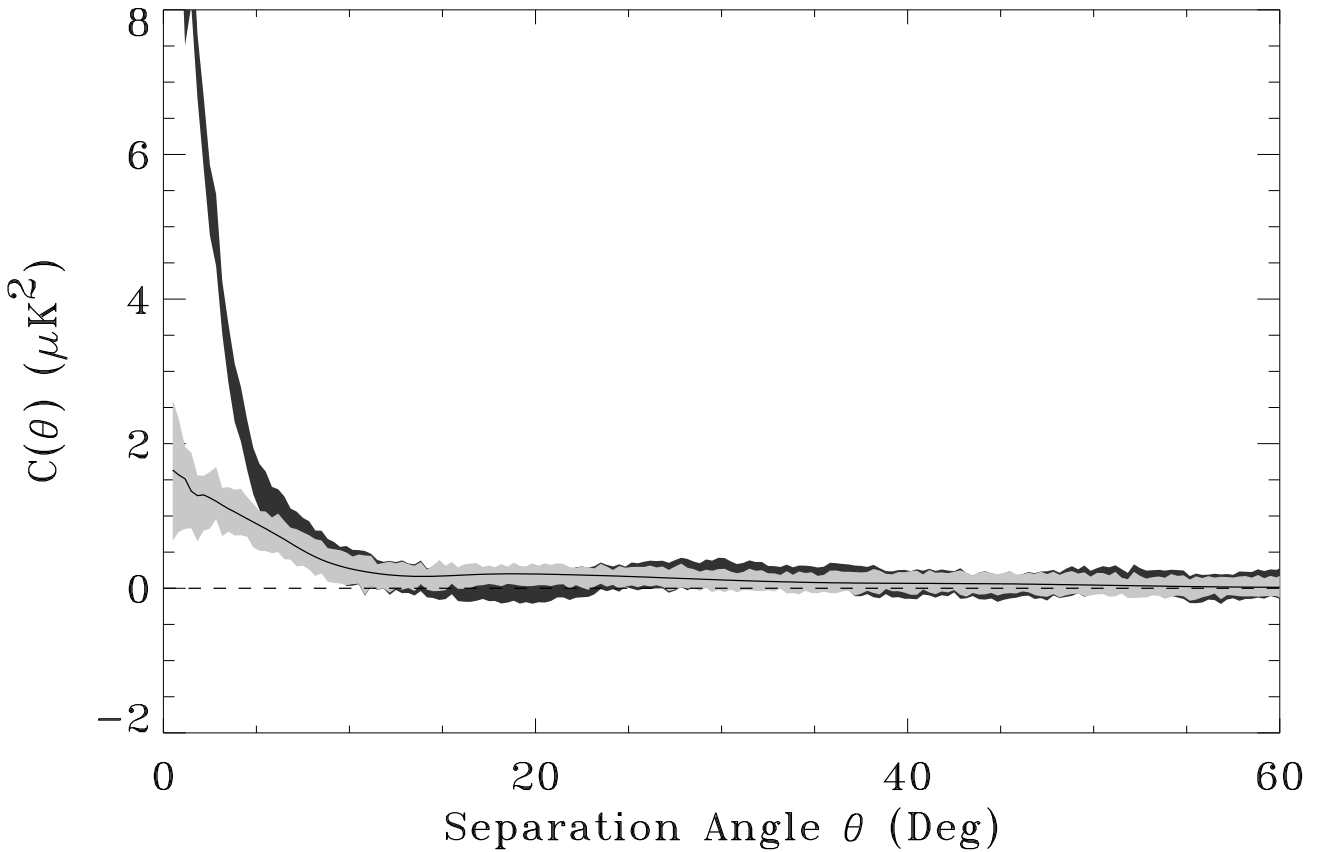,height=5.0in,angle=90}}
\caption{\single12 
Auto-correlation function 
of polarized emission (Q component)
for a reionized CDM model.
The gray bands show the 68\% confidence range in each angular bin
for models with (dark band) and without (light band) foreground polarization.
All models include Sinstrument noise scaled to the MAP 2-year mission.
The solid line shows the CMB input used to generate the simulations.
Polarized Galactic emission completely dominates the CMB and noise
and biases the polarization autocorrelation,
even for small fractional polarization $f \sim 0.1$.}
\label{QQ_corr}
\end{figure}

Each realization of the sky
includes random CMB and noise contributions,
but uses nearly the same Galactic model.
The only changes in the Galactic model
result from changing the polarization templates $\gamma(l,b)$
in each realization
by adding a constant $\gamma_0$ 
drawn from a uniform population $[-\pi, \pi]$.
This effectively changes the mix of Q and U components 
in a given pixel
from realization to realization,
while retaining a fixed correlation 
between Galactic intensity and polarization amplitude.
We repeat the simulations with different values 
of the coherence angle $\theta_c$
to investigate the effect of large-scale coherence
in the foreground polarization angle $\gamma$.

We restrict analysis to the cut sky with latitude $|b| > 30\deg$.
For such partial sky coverage,
the correlation function
\begin{equation}
C(\theta) = \frac{1}{N_{ij}} \sum_{ij} T^\prime_i T^\prime_j,
\label{corr_def}
\end{equation}
is a simple, powerful statistic for analyzing the sky signal,
where the sum is restricted to pixel pairs $\{i,j\}$ 
separated by angle $\theta$.
The Stokes Q and U components depend on 
a specific choice of coordinate system
and are not rotationally invariant.
All sums in Eq. \ref{corr_def}
use rotationally-invariant quantities $T^\prime$, 
defined as
\begin{eqnarray}
 & & Q^\prime = Q \cos(2 \phi) + U \sin(2 \phi) \nonumber \\
 & & U^\prime = U \cos(2 \phi) - Q \sin(2 \phi) \nonumber \\
 & & I^\prime = I,
\label{prime_def}
\end{eqnarray}
where the angle $\phi$ rotates the coordinate system 
at pixel $i$ about the outward-directed normal vector
to put the meridian along the great circle connecting
pixels $i$ and $j$
(Zaldarriaga \& Seljak 1997;
Kamionkowski, Kosowsky, \& Stebbins 1997).

\section{Discussion}

Figure \ref{QQ_corr} shows the QQ auto-correlation function
for 100 realizations of a reionized CMB model
with and without Galactic emission.
The Galactic signal dominates the fainter cosmic signal,
even for models with maximal CMB polarization.
Direct detection of the cosmic Q or U polarization components 
using the individual Q or U polarization maps
requires weaker Galactic polarization, 
higher observing frequency,
and/or detailed modeling such that
\begin{equation}
\frac{ {\rm Q} }{ 3 ~\mu{\rm K} }
~\left( \frac{ \nu }{ \nu_0 } \right)^{\beta}
~\frac{\delta P}{P} < 0.3,
\label{err_eqn}
\end{equation}
where
$\nu_0$ is the effective frequency of the co-added data,
$\beta$ is the spectral index of the foreground polarization,
and $\delta P/P$ represents the fractional uncertainty in 
any galactic correction applied to the polarization maps.

\begin{figure}[b]
\centerline{
\psfig{file=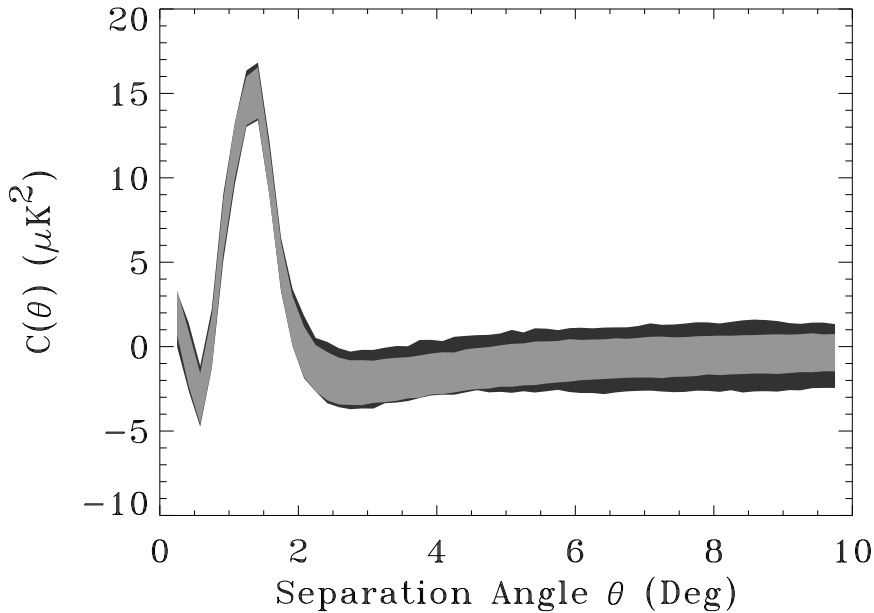,width=5.0in,angle=90}}
\caption{\single12 
Temperature-polarization IQ cross-correlation
for the sCDM model.
The light grey band shows the 68\% confidence range for 100 simulations
of CMB and instrument noise.  
The outer (darker grey) band shows the 68\% confidence range
including polarized Galactic emission 
with coherence angle $\theta_c = 5\deg$.}
\label{IQ_corr}
\end{figure}

The intensity-polarization cross-correlation provides
a more robust probe of cosmic polarization.
Galactic intensity is uncorrelated with CMB intensity.
Even if the amplitude of polarized foregrounds
is correlated with the unpolarized foreground intensity 
(Eq. \ref{pol_def}),
the dependence of the Stokes Q component 
on the polarization angle $\gamma$
makes it equally likely to be positive or negative 
near individual foreground features.
Figure \ref{IQ_corr} shows the mean cosmological signal 
for the IQ cross-correlation in the sCDM model.
With no polarized foregrounds (light gray band),
the cosmic signal at $\theta \approx 1\ddeg3$
is significant at 34 standard deviations
for MAP 2-year noise levels.
When polarized foregrounds are included (dark band),
the significance falls to 31 standard deviations,
still a highly significant detection:
unmodeled Galactic foregrounds 
are unlikely to seriously degrade the IQ cross-correlation.

Polarized foregrounds do not bias the cross-correlation,
but simply introduce an additional noise term.
This additional noise term must be accounted for
in any quantitative analysis.
A commonly used statistic for model fitting is $\chi^2$,
defined as
\begin{equation}
\chi^2 = \sum_{ab} \Delta C_a ~{\bf M}^{-1}_{ab} ~\Delta C_b,
\label{chi_def}
\end{equation}
where
$\Delta C = C_{\rm obs} - \langle C_{\rm model} \rangle $
is the difference between the observed correlation function 
$C_{\rm obs}$
in any given realization
and the mean $\langle C_{\rm model} \rangle$
from some theoretical model,
${\bf M}$ is the corresponding covariance matrix,
and the subscripts $a$ and $b$ refer to angular bins.
We investigate the effects of foregrounds 
by computing $\chi^2$ for simulations with and without polarized signals,
assuming either perfect knowledge 
or no knowledge
of the foreground signal covariance.
Table \ref{chi_table} shows the mean $\chi^2$ (Eq. \ref{chi_def})
for 100 realizations of the simulated IQ cross-correlation function
for the sCDM model and $\theta_c = 5\deg$ 
(the results do not depend strongly on the Galactic coherence angle).
Each row shows the mean $\chi^2$ 
of 100 realizations, 
analyzed using different assumptions for
$\langle C_{\rm model} \rangle$ and ${\bf M}^{-1}$
(computed from the simulations),
corresponding to models with and without polarized emission.
The different rows show the effect of changing the input data set
while keeping model assumptions unchanged.
Different columns show the effect of different model assumptions
for a single data set.

\normalsize
\half12
\begin{table}
\caption{\label{chi_table}Mean $\chi^2$ for Simulated IQ Cross-Correlation}
\begin{center}
\begin{tabular}{l c c c c}
\hline
Data Polarization & \multicolumn{4}{c}{Model Polarization} \\
        & PP$^a$ & PU   & UP    & UU \\
\hline
PP$^a$	& 59	& 92	& 397	& 499 \\
PU	& 61	& 59	& 365	& 433 \\
UP	& 473	& 568	& 59	& 111 \\
UU	& 457	& 537	& 34	& 59 \\
\hline
\end{tabular}
\end{center}
{\single12 $^a$ Letter pairs refer to polarized (P) 
or unpolarized (U) emission
for the CMB and Galaxy, respectively: PU indicates polarized CMB emission
and unpolarized Galactic emission, while UP indicates unpolarized CMB
emission and polarized Galactic emission.
Each $\chi^2$ value uses 59 angular bins.} \\
\end{table}

Polarized Galactic emission
does not prevent a significant detection of cosmological signal.
If the input data contains cosmological polarization
but unpolarized foregrounds (row PU),
the increase in $\chi^2$
when the model assumes no cosmic polarization
allows us to reject this null hypothesis
at 34 standard deviations
(columns UU--PU).
In the likely case that the sky contains 
significant unmodeled foreground polarization,
we would calculate $\chi^2$ using the same model functions as before
(since we lack knowledge of the correct functions including foregrounds)
but with input data now containing 
both cosmic and foreground polarization (row PP).
Two effects are apparent:
we still reject the null hypothesis,
but the high $\chi^2$/DOF indicates
that the data are not well represented by the model.

The first two rows of Table \ref{chi_table}
show that polarized foreground emission
is unlikely to falsify a detection in the IQ cross-correlation
if a cosmic signal exists.
The third row demonstrates that polarized foregrounds 
are also unable to mimic a cosmic signal 
in the event that the CMB is, in fact, unpolarized.
Even if the only signal in the Q or U maps
comes from foreground emission correlated with the foreground intensity,
the shape of the resulting IQ cross-correlation function
in any single realization
is sufficiently different from the expected cosmic signal
to reject a cosmic origin at high statistical confidence.
As before, the poor $\chi^2$ of the UP data fitted to the UU model
indicates that additional unmodelled signals exist in the data;
with such high signal to noise ratio,
the correlation functions in the individual frequency channels
could then be used to help identify the source.

A rotationally-invariant correlation function is simple to compute
and is insensitive details of noise coverage 
or the removal of pixels near the Galactic plane.
However, predictions for CMB polarization are more commonly expressed
in terms of the power spectrum,
which is much more difficult to extract from partial sky maps.
Since the power spectrum is the Legendre transform of the
2-point correlation function,
we may approximate the cut-sky power spectrum
by calculating the Legendre transform of the cut-sky correlation function
for each realization,
then averaging over realizations at each multipole $\ell$.
To speed the Monte Carlo simulations,
we only compute each correlation function 
over the region $\theta < 10\deg$
where the cosmological signal is largest.
This truncation is equivalent to smoothing the power spectrum
with a filter of full width at half maximum $\Delta \ell \approx 20$.
Window effects from the Galactic cut are substantially smaller.

\begin{figure}[b]
\centerline{
\psfig{file=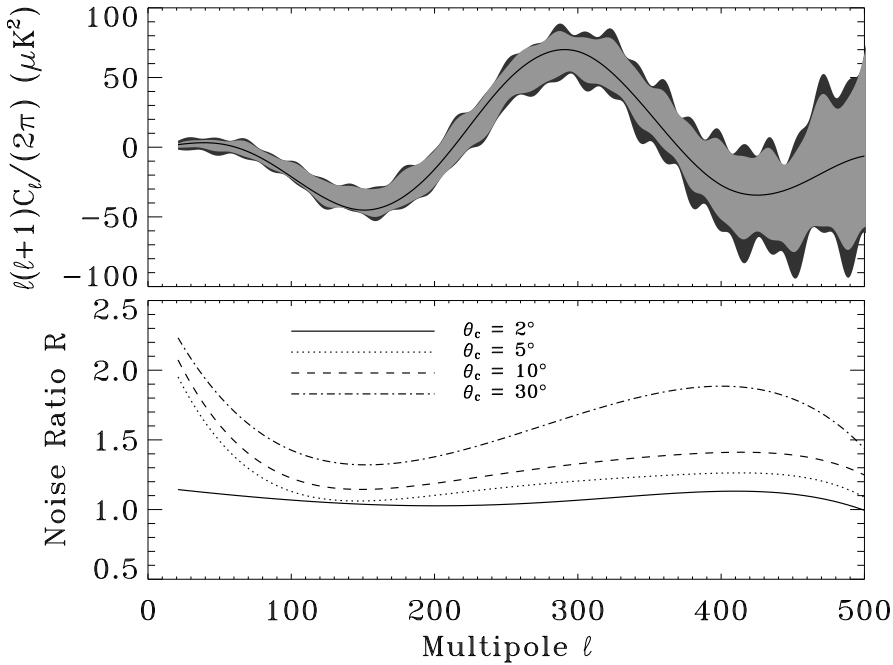,width=5.0in,angle=90}}
\caption{\single12
(top panel) 
Temperature-polarization power spectrum derived 
from the Legendre transform of individual correlation functions.
The dark and light bands show the 68\% confidence range for 100 simulations
with and without polarized foregrounds, respectively.
The solid line indicates the theoretical power spectrum
used to generate the input correlation functions.
(bottom panel)
Ratio $R$ of the scatter at each multipole
for models with foregrounds 
compared to models without foregrounds.
All simulations include CMB and instrument noise.
Polarized foregrounds do not dominate the effective noise. }
\label{cross_power}
\end{figure}

Figure \ref{cross_power}
shows the IQ power spectrum 
from the same sCDM data as Figure \ref{IQ_corr},
along with the theoretical power spectrum 
used to generate the polarized maps.
The simple approximation is seen to be valid
(up to minor ``ringing'' from the Galaxy cut)
and provides a quick representation of the data
in a form readily comparable to theory.
We use this approximation to quantify the relative contributions of
instrument noise and foreground emission to the IQ cross-correlation.
We define
$ R = \delta C_\ell^\prime / \delta C_\ell $
as the ratio of the scatter at each multipole
for models with foregrounds ($\delta C_\ell^\prime$) 
compared to models without foregrounds ($\delta C_\ell$),
where both models include the cosmic signal and instrument noise.
The ratio $R$ is similar to 
the degradation factor DF of Tegmark et al.\ (2000),
except for the foreground cleaning used therein.
It differs from the ``quality factor'' 
of Bouchet et al.\ (1999),
which compares the recovered signal $ C_\ell^\prime$ 
to the true (noiseless) sky signal.
The ratio $R$ shows the effect of the foregrounds {\it only}
and thus separates the question
of foreground removal (to correct or not to correct)
from experiment design (choice of observing frequencies and noise level).
The ratio $R$ is plotted in the bottom panel of Figure \ref{cross_power}
for several values of foreground coherence angle $\theta_c$.

Since polarized foreground emission does not bias the mean cosmological signal,
we can treat it as an additional noise term which adds in quadrature 
with the cosmic variance and instrument noise.  
By fitting the ratio $R$ to models 
with different polarization amplitudes $Q$
and correlation angles $\theta_c$, 
we derive an approximate relation
\begin{equation}
R \approx \left[ 1 + 0.12
\left( \frac{ {\rm Q} }{ 3 ~\mu{\rm K} } \right)^2
\left( \frac{\theta_c}{2\deg} \right)^{0.9}
\right]^{0.5}
\label{scale_ratio}
\end{equation}
for high-latitude data with MAP two-year noise and $\ell < 500$,
where the foreground amplitude Q is evaluated
at the effective frequency of the co-added CMB maps.
Likely polarized foregrounds do not dominate
the noise budget for the temperature-polarization cross-correlation;
foreground removal techniques that increase the effective instrument noise
by more than 30\% are likely to do more harm than good.

\section{Conclusions}

We use simulations to investigate the effect 
of polarized Galactic foreground emission
on the ability of planned CMB missions
to detect and model CMB polarization.
Emission from likely polarized sources
(synchrotron and spinning dust)
would dominate the polarization of the microwave sky
at frequencies below 90 GHz
if the high-latitude Galactic foregrounds are 10\% polarized on average.
Detection of a cosmic signal in the Stokes Q and U parameters
would then require additional effort
(using multi-frequency maps)
to model and separate Galactic from cosmic emission.

Cross-correlation of the temperature and polarization maps
provides a more robust method to detect CMB polarization,
even if the foreground polarization is well correlated 
with the foreground intensity.
Polarized foregrounds twice as bright as the CMB polarization
still allow detection of the CMB signal 
at more than 31 standard deviations
in the IQ cross-correlation
for noise levels appropriate to the forthcoming MAP mission.
The uncertainty in the recovered temperature-polarization power spectrum
is dominated by cosmic variance and instrument noise,
not by the additional ``noise'' caused by foregrounds.
Methods that reduce the foreground signal using linear combinations
of multi-frequency data
can do more harm than good 
if the increase in instrument noise in the corrected data
is larger than the foreground ``noise'' in the un-corrected data.

\vspace{0.5 in}
This work was supported in part by 
the Long Term Space Astrophysics research program
under NASA RTOP 399-20-61-01.

\clearpage
\begin{center}
\large
References
\end{center}

\normalsize
\half12

\refitem
Bennett, C.\ L., et al.\ 1992, ApJL, 396, L7

\refitem
Bennett, C.\ L.,  et al.\ 1995, BAAS, 187, 7109;
http://map.gsfc.nasa.gov/

\refitem
Bouchet, F.\ R., Prunet, S., and Sethi, S.\ K.\ 1999, MNRAS, 302, 663

\refitem
Brouw, W.\ N., and Spoelstra, T.\ A.\ 1976, A\&AS, 26, 129

\refitem
de Oliveira-Costa, A., et al.\ 1998, ApJL, 509, L9

\refitem
de Oliveira-Costa, A., et al.\ 1999, ApJL, 527, L9

\refitem
Draine, B.\ T., and Lazarian, A.\ 1998, ApJL, 494, L19

\refitem
Draine, B.\ T., and Lazarian, A.\ 1999, ApJL, 512, 740

\refitem
Efstathiou, G., and Bond, J.\ R.\ 1999, MNRAS, 304, 75

\refitem
Ferrara, A., \& Dettmar, R.-J.\ 1994, ApJ, 427, 155

\refitem
Haslam,  C.\ G.\ T., et al.\ 1981, A\&A, 100, 209

\refitem
Hinshaw, G., et al.\ 1996, ApJL, 464, L25

\refitem
Hu, W., and White, M.\ 1997, New Astron., 2, 323

\refitem
Kamionkowski, M., Kosowsky, A., and Stebbins, A.\ 1997, PRD, 55, 7368

\refitem
Kinney, W.\ H.\ 1998, PRD, 58, 123506

\refitem
Kogut, A., et al.\ 1996, ApJL, 464, L5

\refitem
Kogut, A.\ 1997, AJ, 114, 1127

\refitem
Leitch, E.\ M.\ et al.\ 1997, ApJL, 486, L23

\refitem
Lubin, P., and Smoot, G.\ 1981, ApJ, 245, 1

\refitem
Mathewson, D.\ S., and Ford, V.\ L.\ 1970, Mem. R. Ast. Soc., 74, 139
	
\refitem
Reach, W.\ T., Franz, B.\ A., Kelsall, T., \& Weiland, J.\ L.\ 1995,
{\it Unveiling the Cosmic Infrared Background}, ed.\ E.\ Dwek, (New York:AIP)

\refitem
Seljak, U., and Zaldarriaga, M.\ 1996, ApJ, 469, 437

\refitem
Tegmark, M., Eisenstein, D.\ J., Hu, W., and de Oliveira-Costa, A.\
2000, ApJ, in press (preprint astro-ph/9905257)

\refitem
Zaldarriaga, M., Spergel, D., and Seljak, U.\ 1997, ApJ, 488, 1

\refitem
Zaldarriaga, M., and Seljak, U.\ 1997, PRD, 55, 1830

\end{document}